# A Text-to-Speech Pipeline, Evaluation Methodology, and Initial Fine-tuning Results for Child Speech Synthesis.

**Rishabh Jain[1], Mariam Yiwere[1], Dan Bigioi[1], Peter Corcoran[1] (Fellow, IEEE), and Horia Cucu[2]**

[1]School of Electrical and Electronics Engineering, National University of Ireland Galway, Galway, H91 TK33 Ireland
[2]Speech and Dialogue Research Laboratory, University Politehnica of Bucharest, Romania

Corresponding author: Rishabh Jain (e-mail: rishabh.jain@nuigalway.ie).

This research is conducted under the "Data-center Audio/Visual Intelligence on-Device (DAVID)" project, 2020 – 2023, which is funded under the Disruptive Technologies Innovation Fund (DTIF), established under Project Ireland 2040, run by the Department of Enterprise, Trade, and Employment with administrative support from Enterprise Ireland.

**ABSTRACT** Speech synthesis has come a long way as current text-to-speech (TTS) models can now generate natural human-sounding speech. However, most of the TTS research focuses on using adult speech data and there has been very limited work done on child speech synthesis. This study developed and validated a training pipeline for fine-tuning state-of-the-art (SOTA) neural TTS models using child speech datasets. This approach adopts a multi-speaker TTS retuning workflow to provide a transfer-learning pipeline. A publicly available child speech dataset was cleaned to provide a smaller subset of approximately 19 hours, which formed the basis of our fine-tuning experiments. Both subjective and objective evaluations were performed using a pretrained MOSNet for objective evaluation and a novel subjective framework for mean opinion score (MOS) evaluations. Subjective evaluations achieved the MOS of 3.95 for speech intelligibility, 3.89 for voice naturalness, and 3.96 for voice consistency. Objective evaluation using a pretrained MOSNet showed a strong correlation between real and synthetic child voices. Speaker similarity was also verified by calculating the cosine similarity between the embeddings of utterances. An automatic speech recognition (ASR) model is also used to provide a word error rate (WER) comparison between the real and synthetic child voices. The final trained TTS model was able to synthesize child-like speech from reference audio samples as short as 5 seconds.

## I. INTRODUCTION

The bulk of recent research into human speech has focused on neural network techniques to improve speech understanding and recognition or to provide simplified, high-quality text-to-speech (TTS) models that can directly convert written text into natural speech. The most highly developed domain for such research has a focus on spoken English and is based on native-speaker adult voice data samples. Automated speech recognition (ASR) is a core element of modern consumer technology user interfaces employed in smart-speaker and voice command interfaces. For interactive chatbot and voice services, TTS models are also important, and the most advanced models can incorporate emotional and prosodic elements into the generated speech output.

More recent research into low-resource languages and other low-resource aspects of human speech, such as accented and prosody-aligned speech has started to see improvements for both ASR and TTS [1]. Another aspect of human speech of growing importance is that of child speech. Child speech differs significantly from those of adult speech, falling into a narrow range of variation and with higher pitch levels. Furthermore, children's speech patterns are more inarticulate and can vary widely in terms of volume, pacing, and emotional expressivity. These challenges are further amplified by the relatively small number of public child speech corpora that are available with useful annotations.

Current work done on TTS for child's voices is limited. This is mainly due to the lack of child voice datasets and difficulty in creating such datasets. As TTS models require hundreds of hours of annotated data for training [2], performing TTS for child voices can be quite challenging. The focus of this work is to explore the potential of state-of-the-art (SOTA) TTS to build a pipeline for the synthesis of children's voices with low data requirements. More specifically, if we can build such a pipeline and demonstrate that it can reliably synthesize a useful number of distinct children's voices, this pipeline would enable the creation of large synthetic datasets that could further improve other aspects of child speech research such as automatic speech recognition (ASR), speaker recognition, etc. To better elaborate on this hypothesis, it is useful to review current SOTA in TTS technologies, followed by a similar consideration for review in child speech research.

### A. RELATED RESEARCH IN TTS

Early research work on TTS synthesis can be traced back to four/five decades ago when the task of TTS was commonly tackled using concatenative and parametric approaches [3]–[7]. Although these early methods were successful in generating speech from text, they generally lacked naturalness. The audio generated using these approaches was kind of muffled and sounded very robotic.

Recent state-of-the-art TTS models are largely based on deep neural networks (DNN) and can achieve more natural-sounding/human-like synthesized speech. With the introduction of Tacotron [8], a neural sequence to sequence the TTS model, the quality of speech synthesis improved significantly. While there are newer approaches that are more efficient or use smaller models, etc., it is still representative of SOTA for the quality of the synthesized speech and is used as a benchmark for comparison with newer methods. Nonetheless, Tacotron TTS is not very robust as it sometimes skips certain words and it also suffers from low inference speed [9]. Several methods have since been proposed to improve upon it such as Tacotron2 [10], FastSpeech [11], FastSpeech2 [12], Transformer TTS [13], FlowTTS [14], GlowTTS [15], etc. Similarly, there have been several improvements over the quality of synthesized waveforms by the introduction of SOTA Vocoders such as WaveNet [16], WaveGlow [17], MelGAN [18], Hifi-Gan [19], WaveRNN [20], etc. These TTS models supported single speaker synthesis, but Deepvoice2 [21], introduced the use of speaker verification models [22]–[25] to achieve Multi-speaker TTS [26]–[34].

### B. CHILD SPEECH – LITERATURE AND CHALLENGES

While all SOTA TTS systems rely on large datasets to train, the datasets mostly comprise speech taken from adult native English speakers; hence, for low-resource languages and other target groups such as non-native adult speakers and child speakers, there remain challenges developing effective and suitable TTS models. Specifically, in comparison with adult TTS, child TTS has gained very little to no attention from the TTS research community. With the current trend of data-hungry DNN-based TTS, TTS for children has practically been neglected due to the lack of large publicly available children's speech datasets for training such networks. Prior to this DNN era, researchers worked on TTS for children using HMM-based models [3], [6].

Collecting data for child speech research can be a challenging task. Most TTS datasets are created in studios with expensive equipment: an adult will be using a microphone to create a clean, noiseless, easy to understand, and meaningful audio. This task is not easy to produce and even more difficult to implement with a child.

One of the main differences between adult speech and child speech is the fundamental frequency. The pitch for children is significantly higher than that of an adult [35]–[38]. The pitch for an adult voice lies between 70 to 250 Hz whereas the pitch for the children's speech is between 200 to 500 Hz [39]. There is also a difference in the speaking rate of children. It was noticed that average phoneme duration is longer in children, therefore, leading to longer speaking rates as compared to adult speech [38], [40]–[42]. The vocal tract of an adult is larger as compared to children's vocal tract and therefore produces different prosody features as compared to an adult voice [43], [44]. Hence, a substantial difference in children's voice characteristics and features can be seen as compared to an adult voice.

Our work aims to solve the problem of TTS for children using DNNs. To solve this problem, the huge challenge of limited publicly available children's speech datasets must first be overcome. To this end, this study considered the use of an existing multi-speaker children's speech dataset [45], which comes with an incomplete set of utterance transcriptions. In addition, this dataset has a lot of unusable data, such as empty/blank entries, extremely long entries as well as inaccurate transcriptions. Firstly, the dataset is cleaned up to create a subset that is suitable for training a neural TTS model. Secondly, with the cleaned-up dataset, a multi-speaker TTS model is trained to generate synthetic speech for multiple child speakers as a proof of concept for children's TTS. The training involved fine-tuning an existing adult multi-speaker TTS model [33] by way of transfer learning, with a few modifications as explained in later sections. This approach involves the training of a separate speaker verification model, and it was preferred because it reduces the problem at hand in two ways:

1) To train the speaker verification network, transcriptions for the speech dataset are not required. Only the speaker identities for the utterances are needed and it can also be trained on noisy speech without any negative effects. This means that even the noisy children's speech dataset, which has incomplete transcriptions, can be useful in training the verification model.

2)  Being a transfer learning process, the pretrained TTS model can be finetuned sufficiently using the resulting cleaned set of children's speech data.

Subjective and Objective Evaluation performed on the synthesized child voices confirms that the child voices generated synthetically are very close to the real child voice in terms of different acoustic features and MOS.

The rest of this paper is organized as follows. Section II describes the methodology and datasets used in this study. The experiments are presented in Section III, the result and evaluation in Section IV, and finally, the conclusion and future work in Section V.

## II.  PROPOSED METHODOLOGY

### A. DATASETS USED IN THIS STUDY
The nature of this study, considering the challenge of limited children's speech datasets and the multi-step training process involved, calls for the use of multiple large datasets, including adult speech datasets. All these datasets are described in Table 1.

**TABLE 1: DATASET USED IN THIS WORK**

| Dataset | # of speakers | # of hours | # of utterances |
|---|---|---|---|
| MyST | 1371 | 393 | 228,874 |
| VoxCeleb1 | 1251 | 352 | 153,516 |
| LibriSpeech | 2484 | 1000 | - |
| VCTK | 110 | 44 | 400 each |

- **MyST**[45]: My Science Tutor (MyST) children's corpus consists of child speech collected using the interaction of the student with a virtual science tutor. The data consists of 393 hours of child speech collected from 1371 students producing a total of 228,874 utterances. 45% of the data is transcribed at word-level leading to about 103,082 utterances and around 208 hours presented in a .trn file format. The MyST corpus is used for this paper because it is the biggest corpus of child speech freely available for research use.

- **VoxCeleb1**[46]: VoxCeleb 1 contains audio recordings of celebrity voices extracted from YouTube. It contains 153,516 utterances from 1,251 speakers.

- **LibriSpeech**[47]: LibriSpeech is a read English speech dataset derived from audiobooks. The data contains approximately 1000 hours of adult speech data from 2400 speakers. The data is divided into two sets, "clean" and "other" where the clean set contains less noisy data as compared to the other set. The "clean" set contains 460 hours of data, and the "other" set contains 540 hours of data.

- **VCTK**[48]: This dataset contains speech recordings from 110 English speakers each reading about 400

sentences from a newspaper. The data contains recordings from various English accents and is highly used in multi-speaker TTS research.

### 1)  PROBLEMS IDENTIFIED IN MYST DATASET
A study on the MyST dataset was performed to measure the amount of data in MyST with and without a transcript. This was done to extract data available with annotation and to see if it can be used for training TTS. A comparison between the complete MyST dataset and filtered MyST dataset where transcripts are available is presented in Table 2. This table provides information on the utterance count and duration of utterance concerning the duration range.

**TABLE 2: MYST DATASET COMPARISON [COMPLETE VS WITH TRANSCRIPT]**

| | MyST (Complete) | | MyST (with transcripts) | |
|---|---|---|---|---|
| Seconds (range) | # of utterances | Duration (in hours) | # of utterances | Duration (in hours) |
| 0-5 | 113,219 | 62.19 | 51,350 | 27.98 |
| 5-10 | 43,782 | 87.78 | 20,723 | 41.78 |
| **10-15** | **22,321** | **75.80** | **11,096** | **37.78** |
| 15-20 | 11,477 | 54.86 | 6,067 | 29.04 |
| 20-30 | 9,282 | 61.89 | 4,991 | 33.28 |
| 30-40 | 2,796 | 26.50 | 1,542 | 14.61 |
| 40-50 | 930 | 11.40 | 517 | 6.32 |
| 50-60 | 347 | 5.24 | 184 | 2.78 |
| 60-70 | 146 | 2.61 | 83 | 1.48 |
| 70-80 | 74 | 1.52 | 38 | 0.78 |
| 80-90 | 53 | 1.25 | 23 | 0.54 |
| 90-100 | 19 | 0.49 | 5 | 0.13 |
| 100 Above | 41 | 1.95 | 17 | 0.94 |
| **Total** | **228,874** | **393.51** | **103,082** | **197.48** |

From Table 2, it was observed that 197.5 hours of child speech data is available with annotation. Although a lot of this data can't be used having different memory requirements on different GPUs. In our experiments, that data between the range of 10-15 seconds to be most useful.

Some initial experiments were performed on the MyST dataset without using the Multi-speaker TTS approach (see section III.A). The results obtained from these experiments were unintelligible. The output waveforms did not have any phonetic meaning and were missing quite some pronunciations. On a more detailed manual inspection of the MyST dataset, a few common problems were identified. The transcripts of some example audio files are listed below to illustrate the problems in the MyST dataset:

- Audio files containing noise in their utterances without any phonetic meaning.
  - "<noise>"
  - "it's glowing <breath>"
- Audio files that are not coherent or indiscernible.
  - "in oxygen right <indiscernible>"
  - "can hear sound because of that <indiscernible>"

- Audio files are too small in length
  - "energy <noise>"
- Audio files are too long
  - "it's trying to show us that all the things that it needs all the things that the plants needs to grow it needs soil on the bottom it needs at least a ground a top the a a top to lay on for the plant to grow so you can see it that's only with flowers and plants it's not with vegetables and it needs and it needs the energy from the sunlight to grow and it needs water because somebody's watering the plant."
- Transcription containing text with no phonetic information.
  - "(()) (()) (())"
- Repetition of words/stammering noticed in children's voices.
  - "um we measured how big a millimeter meter is a meter and a kolome- a *kilometer*"

Our examination of MyST led us to further clean the MyST dataset for TTS training. In this process a subset of MyST, hereafter referred to as TinyMyST was created.

### 2) TINYMYST

It is a small subset of the MyST dataset created using various pre-processing scripts to make the data suitable for TTS acoustic model training. MyST was cleaned to select only audio files with existing transcriptions. All audio files lesser than 10 seconds and greater than 15 seconds were removed. The utterances shorter than 10 seconds contained mostly noise or unintelligible speech and those longer than 15 seconds were removed to avoid GPU memory overflow during training. All the transcript files were converted from *.trn* format to *.txt* file format.

**TABLE 3: MYST VS TINYMYST**

|  | MyST (Complete) | MyST (with transcripts) | TinyMyST (Usable for TTS) |
|---|---|---|---|
| **Speakers** | 1371 | 738 | 670 |
| **Duration (in hrs)** | 393.5 | 197.5 | 19.22 |
| **# of utterances** | 228,874 | 103,082 | 7152 |
| **Mean duration per speaker** | 17.22 mins | 16.05 mins | 1.73 mins |
| **Speaker with most data** | 013023 (110.38 mins) | 013023 (81.71 mins) | 013023 (8.77 mins) |
| **Speaker with least data** | 012002 (0.96 secs) | 007389 (0.96 secs) | 018216 (10.2 secs) |

**mins: minutes, secs: seconds**

The TinyMyST dataset still contains a lot of noisy data. Some of the excessively noisy data were removed manually by inspecting the transcripts and listening to the audio samples. The data obtained after cleaning contained 7152 utterances and accounted for 19.22 hours. A detailed comparison of MyST and TinyMyST datasets was performed to see differences in the two datasets in terms of speaker identities and utterances (see Table 3). TinyMyST dataset on average contained 1.72 minutes per speaker having 670 speakers. Speaker identity '013023' had the most data with 8.77 minutes and speaker identity '018216' has the least data with 10.01 seconds. The speaker '013023' had the most data in MyST as well to be around 110 minutes.

To extract more TTS usable data, an audio sample from more than 15 seconds long can be used to split them into smaller chunks. A forced aligner[1] is used to align the audio files with transcripts. Time alignment information from the alignments to split the longer audio files into smaller samples, however, it was observed that the audio alignment was not very accurate for the child speech and there were a lot of mismatches between the transcripts and audio files. This was probably due to fact that the pretrained forced aligner was trained on adult speech and doesn't work very well for aligning child speech. Therefore, TinyMyST was used (as described earlier) for performing all the child TTS experiments.

### 3) DATA PREPROCESSING FOR TTS USAGE

LibriSpeech and TinyMyST datasets were preprocessed as per the guidelines mentioned in LibriTTS[49]. The LibriTTS dataset was specifically created for TTS research, therefore similar guidelines were followed in our experiments. The following changes were made:

- Audio files were converted to 16-bit depth audio files with 24Khz sampling rate (WAV format), This was done using the pydub[2] audio library.
- Text data was normalized by replacing abbreviations and punctuations.
- Whitespaces were normalized
- All characters were made uppercase.

### B. MULTI-SPEAKER TTS MODEL

The neural network used to achieve TTS for children in this study is based on [33], It works by combining a speaker verification network with the SOTA Tacotron TTS model. Though Tacotron is SOTA for TTS, it was designed to be trained using a single-speaker speech dataset such as the LJSpeech [50] dataset, hence, it can only synthesize speech with acoustic characteristics of the single speaker whose data was used in training. To function effectively for multiple speakers, Tacotron needs to be adapted for that purpose. This adaptation has been achieved in this multi-speaker TTS model [33] by introducing different speaker identities in the form of speaker embeddings as additional input to the Tacotron network. As a result, the multi-speaker TTS [33] comprises three different neural network models, each of which focuses on a specific subtask namely, Speaker Encoder used for speaker verification task, Acoustic model used for spectrogram synthesis, and a Vocoder for audio waveform generation (as shown in Figure 1).

---

[1] https://github.com/MontrealCorpusTools/Montreal-Forced-Aligner
[2] https://github.com/jiaaro/pydub

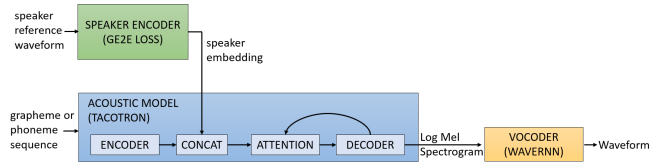

**FIGURE 1.** Model Overview: Speaker Encoder, Acoustic Model, and Vocoder Models trained independently (from [33]).

For our work, generalized end-to-end (GE2E) loss was used for speaker verification [22], Tacotron1 as an acoustic model [8], and WaveRNN as Vocoder [20]. The original approach [33] is adapted for child speech synthesis by first pretraining the model on an adult speech dataset after which, it is fine-tuned with the child speech dataset.

The speaker encoder generates speaker embeddings, encoding speaker identity information extracted from the utterances. Similar voices are mapped closer to each other in a latent space representation. The acoustic model generates spectrograms from text conditioned on the speaker embeddings. The vocoder then converts these spectrograms into audio waveforms. At inference time, a short reference utterance (ground truth) of a child's voice is passed through the speaker encoder to generate the corresponding speaker embeddings, on which the acoustic model will be conditioned. The three different neural network models are described as follows.

### 1) SPEAKER ENCODER
The first stage of the multi-speaker TTS training involves the training of a speaker verification (speaker encoder) model. Speaker Verification is the process of determining if an utterance belongs to a specific speaker. The speaker encoder is used to train the model for the speaker verification task using a mix of noisy and clean speech data without transcripts. The data used consists of both adult and child speech data from thousands of speakers (see Table 4). This was done to introduce both child and adult speakers in the model for better generalization. The output of this model conditions the acoustic model to generate the required mel-spectrograms from a reference speech signal of the target speaker. The model is trained to capture the characteristic features of different speakers.

The model takes input as log mel-spectrograms computed from utterances of each speaker, trains using the GE2E loss and converts them into a fixed dimensional vector called d-vectors. These d-vectors are optimized over GE2E loss to differentiate the speakers, such that the same speakers have embeddings with high cosine similarity and different speakers are far apart in the embedding space.

During training, complete utterances are segmented into partial utterances of 1.6 seconds. These parameters were kept the same as explained by authors [51][22]. The utterance embedding is calculated using 800ms windows for inference, with a 50% overlap. The silence was removed from the utterances using the webrtcvad[3] tool for Voice Activity Detection (VAD). Each segment is passed through the network individually, the outputs are averaged and normalized to create the final utterance embedding as described in [22].

The encoder model is trained using 4 datasets, MyST, VoxCeleb1, LibriSpeech, and VCTK. Equal Error Rate (EER) is used as a metric for the validation of the speaker encoder. The default EER metric from [51] is used in this work as authors of [33] have not explicitly specified the training, test, and validation criterion they are using for EER calculation. The EER values are presented in Table 4. The model trained for one million steps was used in the multi-speaker TTS model as relatively insignificant improvements were seen in the EER after this point.

**TABLE 4: SPEAKER ENCODER TRAINING DETAILS**

| Dataset used for Encoder Training | Size (in hours) | Iterations | EER |
|---|---|---|---|
| - MyST<br>- VCTK<br>- VoxCeleb1<br>- LibriSpeech [Other] | 1329 | 1M | 5% |

All the datasets were pre-processed into the coding format required for training the encoder as described in [51]. Even though half the MyST dataset is not transcribed, the complete MyST dataset can be used for Speaker Encoder training as it does not require any transcription data. The pipeline for the speaker encoder training can be seen in Figure 2.

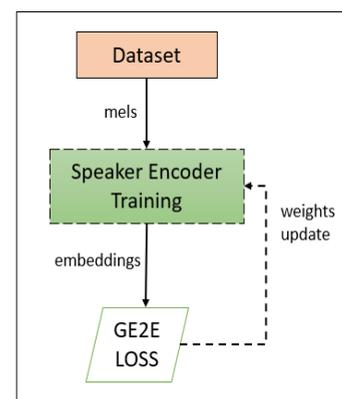

**FIGURE 2.** Pipeline for Speaker Encoder training. The dotted line represents the training loop for the Speaker Encoder training.

A UMAP projection [52] is created to visualize the training by taking a random set of 10 utterances from 10 speakers. Utterances with similar embeddings are located close to each other in the latent space representation and have similar speaker characteristics.

---

[3] https://github.com/wiseman/py-webrtcvad

This model creates individual clusters of speaker embeddings as can be seen in the UMAP projection (see Figure 3). Each point on UMAP represents an utterance. The same color points represent the same speaker. Encoder gradually learns to separate the speakers. Initially, there is a lot of overlap across speakers, but eventually, each speaker has their utterances clustered and well separated from the other speakers. The training evolves with increased training steps.

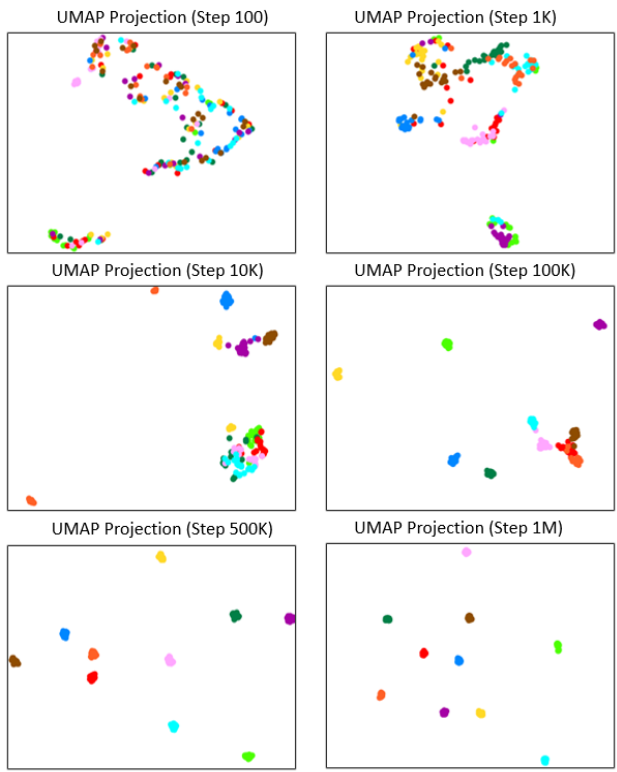

**FIGURE 3.** UMAP projections at different training steps for speaker encoder training. Ten different colors represent ten different speakers with ten utterances each.

### 2) TACOTRON ACOUSTIC MODEL

For the speech spectrogram synthesis, the TTS model architecture and hyperparameters used in this study are the same as in the work of [51] (More details are provided in Section III). The authors used a modified version of the original Tacotron architecture [8]. The model consists of an encoder, an attention-based decoder, and a post-processing network. Since Tacotron is originally a single-speaker TTS model, it was modified to work for multi-speaker TTS by connecting the speaker encoder to it. Speaker embeddings from the encoder are concatenated with text (character/phoneme) embeddings from the text encoder, after which an attention mechanism is applied prior to decoding into a spectrogram. Unlike the speaker encoder, the acoustic model takes in both audio(utterance) and associated text(transcript) as inputs.

In this work, the acoustic model was first trained with only adult speech data (acoustic model training I), specifically, the Librispeech 'clean' data, until it started to converge at 250k steps and then finetuned with the TinyMyST child speech dataset (acoustic model training II) for up to 750k additional steps (more details in Section III). The pipeline for the acoustic model training can be seen in Figure 4.

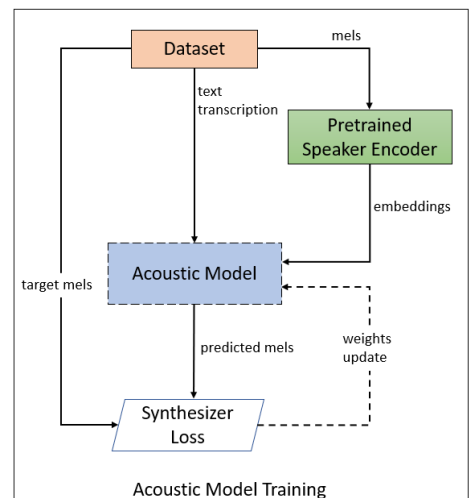

**FIGURE 4a.** Pipeline for Acoustic Model training. A model with solid contour represents the pretrained model. Dotted contours represent the acoustic model training loop.

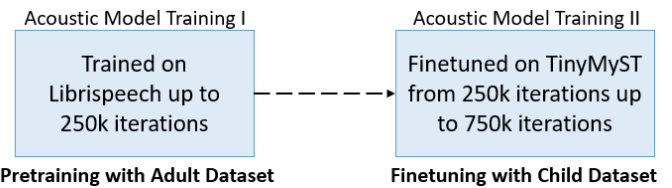

**FIGURE 4b.** Fine-tuning step for Acoustic model training. Acoustic model training I represent the acoustic model being trained with LibriSpeech dataset for up to 250k iterations. Acoustic model training II represents fine-tuning the acoustic model I with the TinyMyST dataset from 250k iteration onwards up to 750k iteration.

### 3) WAVERNN VOCODER

The vocoder used is WaveRNN[20], which is an improvement over the WaveNet[16] originally used by the authors of [33]. WaveRNN is a recurrent network for performing sequential modeling of audio from mel-spectrograms. An alternative version of WaveRNN is used, having a few architectural changes as provided by the author in [53] due to the popularity of the model as it reduces sampling time while maintaining high output quality. WaveRNN uses Gated Recurrent Unit (GRU) in comparison to convolutions used in WaveNet. The input mel-spectrograms and their corresponding waveforms are segmented at each timestamp. A 1D Resnet-like model is used to generate features for layered connections in the alternative WaveRNN architecture. The upsampling is also performed on the mel-spectrogram to match the length of the target waveform. The resulting vector is passed through a combination of GRU and dense layer transformations in four-way connections. These connections are concatenated at different steps to generate the corresponding vector representation. This vector is passed through two dense layer connections which finally generate the encoding of raw audio. The output audio is generated at a 16-bit depth and 16 khz sampling rate.

The predicted mels from the acoustic model trained on LibriSpeech (from acoustic model training I) were used to train the vocoder. The vocoder trained up to 250k iterations was used to generate all waveforms in this study. The pipeline for vocoder training can be seen in Figure 5. Fine-tuning experiments with the TinyMyST dataset didn't improve the quality of the vocoder (more discussion in Future work).

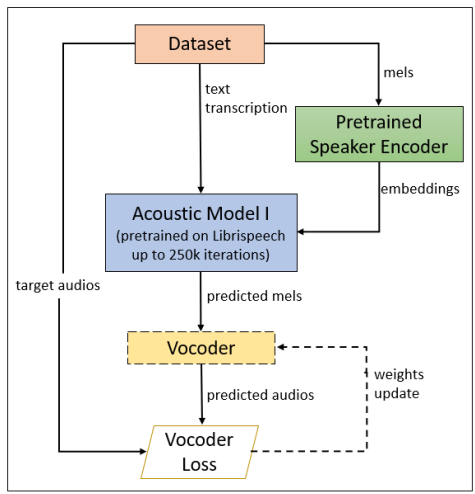

**FIGURE 5.** Pipeline for Vocoder training. Models with solid contours are pretrained models. The dotted contour represents the training loop for Vocoder.

Vocoder for child TTS hasn't been explored before in detail. This is a new area of research. It was observed that WaveRNN has popularly been used as a universal vocoder [54]–[56] and it evidently works well with unseen speakers in multi-speaker models as well [57]. Therefore, for the scope of this paper, WaveRNN (trained on LibriSpeech) is used as a universal vocoder with synthetic child voices.

## III. EXPERIMENTS

### A. INITIAL EXPERIMENTS

In our initial experiments, multiple SOTA TTS models [10], [13]–[15], [21] were unsuccessfully trained, including Tacotron 2[4], using the transcribed subset of the MyST dataset. Figure 6 shows an example of an alignment plot from Tacotron 2 training. As can be seen, there was no sign of alignment even after 200k iterations.

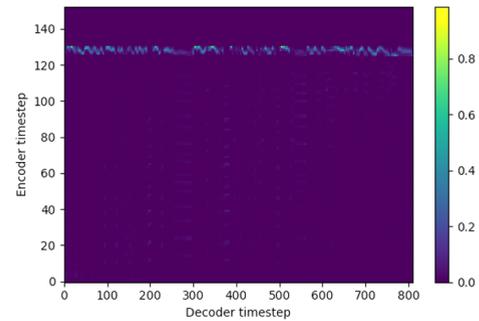

**FIGURE 6.** Alignment plot for Tacotron 2 trained with MyST dataset for up to 200k steps.

Further experiments were conducted using the cleaned subset of MyST (TinyMyST), which showed some alignments as seen in the Tacotron 2 alignment plot in Figure 7. However, though child-like in terms of pitch, the synthesized speech signals were completely unintelligible. Missing information such as 'End of sentence' was observed which mostly contained noise content.

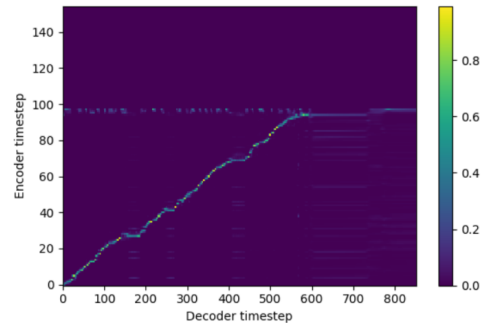

**FIGURE 7.** Alignment plot for Tacotron 2 trained up to 200k steps with TinyMyST Dataset

Next, fine-tuning the pretrained NVIDIA Tacotron 2 model on a single child's MyST dataset utterances resulted in slightly intelligible but highly robotic and unnatural synthesized speech. Figure 8 shows the improved alignment plot from the finetuned Tacotron2.

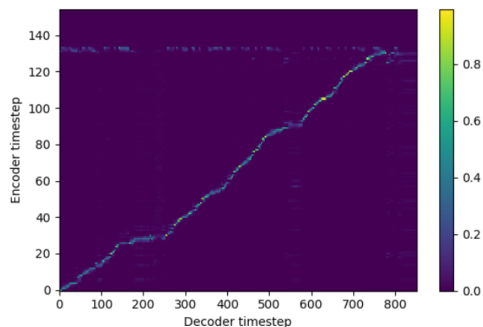

**FIGURE 8.** Alignment plot for Tacotron 2 trained up to 200k steps with TinyMyST Dataset, pretrained with LJ Speech Dataset up to 100k steps.

Since there were not enough MyST utterances for a single child to sufficiently train Tacotron2, different multi-speaker TTS models were explored [21], [26], [27], [58] and the speaker verification-based method [33] produced the most promising results. Hence, this method was used in our main experiments.

---

[4] NVIDIA/tacotron2: https://github.com/NVIDIA/tacotron2

## B. MAIN EXPERIMENTS

As seen in our methodology (Section II.B), a modified approach based on [33] was used by incorporating an extra layer of fine-tuning in the training step.

The proposed neural child voice TTS was trained on a Tesla V100 GPU. Each of the three networks – Speaker Encoder, Acoustic model, and Vocoder were trained separately.

The Speaker Encoder was trained with a batch size of 128 and a learning rate of 0.0001. The model was trained for 15 days for up to 1M steps. EER of 5% was observed at this point with no further improvement afterward. Additional parameters settings are mentioned here[5]. The default embedding size of 256 was used for this training.

For the acoustic model, the network was trained using a learning rate of 0.0001 for 250K steps (pretraining) and 0.00001 for 750k steps (fine-tuning). The batch size was kept constant at 72. Entire training (up to 750k steps) took 9 days to complete. Additional parameters details were kept the same as Tacotron 1, these details are mentioned here[6].

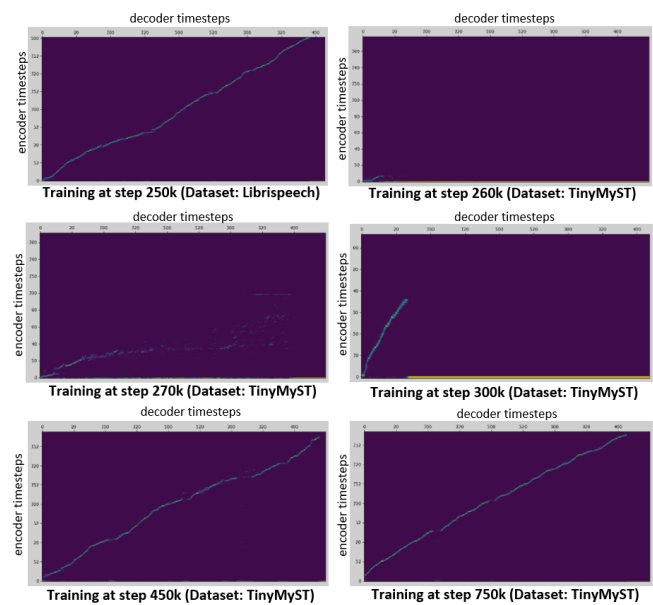

**FIGURE 9.** Alignment plots at different training steps during transfer learning from adult to child Tacotron TTS.

The alignments plot for encoder-decoder timestamps can be seen in Figure 9, the x-axis represents the encoder timesteps and the y-axis represents the decoder timesteps of Tacotron training. The training is done on LibriSpeech up to 250k steps generated a good alignment plot. Alignment weakens when switched to TinyMyST Dataset, but it gradually improves with increasing training steps. During inference, our model was tested on multiple checkpoints taken at intervals of 50k iterations. A few of these iterations are mentioned in Figure 9. Even though the alignment at some of these steps looks the same, an improvement was noted over time with the synthesized child voices. This was determined subjectively during training by listening to the synthetic child speech generated. The training was halted at 750k steps as improvements in the alignment graph had become imperceptible after 700k steps. The output waveform did not show any improvements beyond this step. The model trained up to 750k iterations is used to provide audio samples in this paper.

The Vocoder was trained at a batch size of 128 and learning rate of 0.0001 and took 4 days of training to reach 250k iterations. Most of the parameters for the vocoder were kept the same as the original code[7].

The synthetic child voices during inference were natural sounding and the trained model demonstrated an ability to synthesize quite challenging phrases that were unseen in the TinyMyST dataset. This was tested by using 'tongue twisters' as a reference text for synthesizing speech. However, it was also noted that some phonemes were not synthesized correctly and lost their meaning during synthesis. These findings are discussed in more detail in Section IV.

Code-related material and synthesized speech from these experiments will be made available in our GitHub Repository[8].

## IV. RESULTS AND EVALUATION

The evaluation in TTS is usually done by taking a Mean Opinion Score (MOS) [59] on the synthetic speech for Speech Similarity and Speech Naturalness.

There are many objective and subjective evaluation methods proposed by researchers [60]–[66]. These traditional speech evaluation methods work well for evaluating adult speech but are not so suitable for child speech. A perfect adult speech will contain fluent pronunciation of a word/phoneme however this is not the case for most child speech. Naturalness in child speech includes pauses, breaks, and pronunciation difficulties in the speech. Other challenges were noted with the start and end of phrases where children tend to be somewhat hesitant when starting a phrase and may wander towards the end of one. Children can also mispronounce words, or struggle with the phonetics of a particular phrase.

---

These characteristics tend to manifest in the speech model and a range of artifacts were noted that affect the quality of the phrases synthesized by our pipeline. It was noted that the first or last words in many phrases were either missed entirely or subject to various distortions or artifacts. In the middle of a phrase, there could occasionally be slurring or arbitrary elongating of one or more words. Another artifact observed was that the pace or tone of voice could change abruptly in the middle of a phrase. Despite these artifacts, the majority of phrases were quite intelligible, and a large proportion was also very natural sounding. Therefore, there is a need for a better subjective evaluation method for child speech synthesis.

In the following, we present the results obtained using the proposed subjective evaluation method and the various listening tests performed (subsection 4.A), two objective evaluation methods, based on MOSNet (subsection 4.B) and an ASR system (subsection 4.D) and we evaluate the similarity of the synthesized speech and natural child speech (in subsection 4.C).

### A. PROPOSED SUBJECTIVE EVALUATION METHOD
To check the phonetic coverage of our child speech TTS, Harvard sentences [60] were used, which are a set of 720 phonetically balanced sentences. These sentences cover most of the phoneme range and were designed to be implemented with Voice over Internet Protocol (VoIP) technology. These texts were used to generate synthetic child speech. This was done to check the subjective quality of synthesized audio with respect to phoneme coverage.

Our evaluation method uses a MOS-like evaluation with different categories for scoring. When generating synthetic voices using Harvard sentences, it was observed that some sets of phonemes were not pronounced correctly even when synthesized using different reference child speakers (more detail in a later section). After our initial subjective study of these 720 synthesized audio samples, it was decided that a more detailed evaluation protocol was required to address the various artifacts observed and identify what additional data samples might be needed to further improve our model. For this reason, an evaluation was performed in two phases. For each of the two phases, different evaluators were gathered to perform the speech evaluation. Each evaluator was asked to listen to synthetic audio files using Headphones/Earphones in a noise-free environment. They were asked to rate each of the synthetic voices assigned to them from a range of 1 to 5, for each of the different categories in two phases. The categories included Speech Intelligibility, Voice Naturalness, and Voice Consistency. Voice Consistency contained three sub-categories of its own namely, Start of Phrase quality, Middle of Phrase quality, and End of Phrase Quality.

Evaluation data was provided in a OneDrive Environment. All the synthetic voices were shared in a common OneDrive folder to the evaluators and a common spreadsheet was circulated containing the utterance ID of Harvard sentences used for synthesizing a child's voice. While listening to many different natural child voices, it was also noticed that recorded child audio can be a difficult task to understand if not provided with a suitable transcript. Some of the child's speech can be non-meaningful as mentioned in problems with the MyST section. After performing many different tests and trials using child speech, the use of transcripts as a part of MOS-based evaluation is considered to be a more natural way of evaluating child speech. Therefore, corresponding transcript information is also provided in the spreadsheet to each evaluator to base their conclusion on 'what they hear in child audio' and 'what they read in child transcripts'. This way more coherent patterns can be observed among the phonemes and graphemes in a child's voice for each of the mentioned categories. An example of this spreadsheet can be seen here[9].

By performing the evaluation using OneDrive environment, it was easy to distribute the synthetic speech files to different evaluators without having to spend time and resources on expensive Mushra-based evaluations [67] or crowdsourcing the evaluation task on platforms like Amazon Mechanical Turk (AMT) [59], [68]. Mushra-based evaluations were also avoided due to potential biases that can occur in these tests and how these biases can impact synthetic child voice evaluation for MOS [69]. Most of these TTS evaluations have been conducted before with synthetic adult speech, this novel synthetic evaluation is implemented for first-time with synthetic child speech. Using a common spreadsheet made it effective to perform analysis of spreadsheet for MOS using pandas and other python-based tools.

#### 1) PHASE-I EVALUATION
For the phase-I evaluation, all 720 Harvard sentences were generated using our proposed TTS method. Two random reference utterances were selected from the TinyMyST dataset and were used to generate all the Harvard sentences. These 720 sentences were shared among 5 evaluators in a spreadsheet document, who rated the voices from 1 to 5 based on **Speech Intelligibility** and **Voice Naturalness**. The MOS ratings from 1 to 5 were further explained in the spreadsheet file as can be seen in Table 5.

**TABLE 5: MOS (FROM 1 TO 5) EXPLAINED FOR SPEECH INTELLIGIBILITY AND VOICE NATURALNESS**

| Score | Speech Intelligibility | Voice Naturalness |
|---|---|---|
| (5) | Voice is clear, all words identifiable | Voice consistently paced with similar timbre across the entire phrase; good voice quality |
| (4) | Voice is mostly clear; single word unclear | Some disjointness in terms of pacing/timbre; mediocre but |

| | | plausible voice quality | |
|---|---|---|---|
| (3) | Voice understandable; multiple words unclear | Significant differences in pacing/timbre across different portions of the phrase; weak voice consistency across different parts of the phrase | |
| (2) | Difficult to understand most words in a phrase | Substantial differences in pacing/timbre across different portions of the phrase; distinctly different voices for different words/parts of the phrase | |
| (1) | Difficult to understand any words in a phrase | No consistency in terms of voice or pacing across the phrase | |

The spreadsheet was later analyzed to get the final mean opinion score in each category. MOS of 3.88 for voice naturalness and 4.13 for speech intelligibility was observed as seen in Table 6.

**TABLE 6: MOS FROM PHASE-I EVALUATION WITH 95% CONFIDENCE INTERVAL**

| Categories | MOS |
|---|---|
| Voice Naturalness | 3.88±0.27 |
| Speech Intelligibility | 4.13±0.34 |

An average score for each of the 720 sentences was calculated for the combined value of speech intelligibility and voice naturalness. All the 720 sentences were sorted into difficult and easy sentences with respect to the children's linguistic capabilities. This was done to keep track of Harvard sentences where synthesized speech becomes unintelligible and inarticulate.

*2) PHASE-II EVALUATION*

After our phase-I evaluation, a common set of sentences were observed where pronunciation sounds unintelligible at the start, middle, or end of sentences for specific words/phonemes. There was an inconsistency in voice quality. These sets of sentences are the ones that were not learned properly during training or were missing in the training dataset for child audio. To make a note of these sentences, extra categories of 'Voice Consistency' were added to the phase-I evaluation. Therefore, all the 3 sub-categories under **Voice Consistency** were used in the second phase of the evaluation. These subcategories included 'Start of Phrase Quality', 'Middle of phrase quality' and 'End of phrase quality'. The MOS ratings from 1 to 5 for each of these categories were also explained in the spreadsheet as mentioned in Table 7.

**TABLE 7: MOS (FROM 1 TO 5) EXPLAINED FOR VOICE CONSISTENCY AND ITS THREE SUB-CATEGORIES**

| Voice Consistency | | | |
|---|---|---|---|
| Score | Start of phrase & first word quality | Middle of phrase & central word quality | End of phrase & last word quality |
| (5) | First word is clear; excellent starting quality & intelligibility | Middle of phrase is clear; excellent voice quality & intelligibility | Last word has excellent quality |
| | Understandable; | Understandable; | Understandable; |

| (4) | minor distortions or noise; low intensity starts of first word | minor distortions or noise; some pacing variations or slurring of single word | minor distortions or noise |
|---|---|---|---|
| (3) | Start of first word unclear or more significant distortions of first work or start of phrase | Significant distortions of middle phrase or slurring of multiple words but still intelligible | Understandable; but significant distortions or noise |
| (2) | First word missing or strongly distorted; distortions or noise impact strongly on phrase intelligibility | Substantial distortions; multi-word slurring or noise impact strongly on phrase intelligibility | Not understandable; substantial distortions, missing or unintelligible words or noise |
| (1) | Start of phrase unintelligible, missing or severely distorted | Middle of phrase unintelligible, missing or severely distorted | Multiple words not understandable |

For the second phase of evaluation, the evaluation was undertaken by 20 evaluators divided into 4 groups. This was done as per the guidelines mentioned in [70] for performing MOS evaluations. For each group, a speaker identity was selected from the TinyMyST dataset. All the speaker identities were sorted, and the top 20 speaker identities were selected, having the most minutes. Among these 20 identities, 4 speaker identities were randomly selected. All the 4 groups are named as '013020', '008045', '002113', and '995737', corresponding to each identity label. This approach was taken to select speakers with the most data and also to keep the process randomized. More information on these selected speakers can be seen in Table 8. This table is also used for speaker similarity and objective intelligibility experiments in the future sections.

**TABLE 8: SELECTED SPEAKER IDENTITY INFORMATION IN TINYMYST VS TTS UTTERANCES FOR THE SAME SPEAKERS**

| Speaker ID | Minutes | Hours | # of Real utterances | # of generated TTS utterances |
|---|---|---|---|---|
| 002113 | 7.75 | 0.13 | 41 | 50 |
| 008045 | 7.88 | 0.13 | 42 | 50 |
| 013020 | 8.77 | 0.15 | 47 | 50 |
| 995737 | 5.70 | 0.10 | 30 | 50 |

A reference child utterance was selected randomly from each of these groups, and 50 Harvard sentences were selected randomly for each of the groups. Therefore, 50 Synthetic utterances were generated, and all the evaluators were asked to rate the utterances assigned to them.

**TABLE 9: MOS FROM PHASE-II EVALUATION WITH 95% CONFIDENCE INTERVAL**

| | SI | VN | VC | | |
|---|---|---|---|---|---|
| | | | SP | MP | EP |
| 013020 | 4.03 | 4.06 | 4.44 | 4.35 | 3.61 |
| 008045 | 3.70 | 3.63 | 3.94 | 3.92 | 3.20 |
| 002113 | 3.52 | 3.49 | 3.62 | 3.89 | 3.18 |

| 995737 | 4.63 | 4.38 | 4.62 | 4.58 | 4.48 |
|---|---|---|---|---|---|
| **Overall MOS** | **3.95±0.30** | **3.89±0.32** | 4.07±0.36 | 4.18±0.21 | 3.62±0.45 |
| | | | | **3.96±0.32** | |

SI: Speech Intelligibility, VN: Voice Naturalness, VC: Voice Consistency, SP: Start of phrase and first word quality, MP: Middle of phrase and central word quality, EP: End of phrase and last word quality.

MOS results from the phase-II evaluation are presented in Table 9. MOS of **3.95** was observed for **Speech Intelligibility**, **3.89** for **Voice Naturalness**, and **3.96** for overall **Voice Consistency** (including the three sub-categories). MOS of 4.07 was observed for 'Start of phrase quality', 4.18 for 'Middle of phrase quality', and 3.62 for 'End of phrase quality'. The MOS score implies that the quality of synthesized child speech is quite good. However, there is still room for improvement in the 'End of phrase quality' of Harvard sentences. There is information loss observed at the end of most sentences containing inarticulate and unintelligent information or noise. The reason for this information loss can be redirected back to the child dataset used for training. Even though TinyMyST is much cleaner than the MyST dataset, it still contains some of the problems seen in Section II.A.1. The information obtained from voice consistency will be discussed more in future work.

A similar experiment was also performed using the real utterances from Table 8 to obtain a baseline MOS on natural child speech. 15 random real utterances were selected from the real speakers mentioned in Table 8. Evaluators were asked to perform a similar evaluation as done in phase-II evaluation for all the selected 60 utterances. A comparison between the baseline MOS on Natural MyST and synthetically generated utterances is mentioned in Table 10.

**TABLE 10. MOS Natural Speech VS MOS Synthetic Speech with 95% Confidence Interval**

| | SI | VN | VC | | |
|---|---|---|---|---|---|
| | | | SP | MP | EP |
| Natural Speech MOS (from MyST) | **4.21±0.42** | **4.05±0.34** | 4.32±0.42 | 4.01±0.6 | 3.9±0.62 |
| | | | **4.08 ± 0.54** | | |
| Synthetic Speech MOS (from Table 9) | **3.95±0.30** | **3.89±0.32** | **3.96 ± 0.32** | | |

SI: Speech Intelligibility, VN: Voice Naturalness, VC: Voice Consistency, SP: Start of phrase and first word quality, MP: Middle of phrase and central word quality, EP: End of phrase and last word quality.

Synthetic Speech MOS for three categories is very close to Natural Speech MOS. There is a **MOS difference** of '**0.26**' for Speech Intelligibility, '**0.16**' for Voice Naturalness, and

'**0.12**' for Voice Consistency between natural and synthetic speech. From Table 10, it can be concluded that the MOS for Natural and Synthetic child speech are quite close to each other. This subjective evaluation approach is proposed as a part of this paper. Due to very limited work done on child speech synthesis, we did not find any reliable way of performing subjective evaluation over the synthesized child speech. From our experience with the evaluation of synthetic child speech, this new metric of evaluation can help evaluate synthetic child speech and can help further this area of research. It is also intended to use this proposed approach for our future work with child speech synthesis.

### B. OBJECTIVE NATURALNESS EVALUATION USING A PRETRAINED MOSNET

For this objective evaluation, a pretrained MOSNet was used, which is trained on VCC 2018 dataset from Blizzard Challenge [66] comprising of adult speech. According to their paper, MOSNet predictions yield a high correlation to human ratings. As MOSNet was trained on adult speech, it is unlikely that it will generalize well for child speech. It won't be possible to train a MOSNet with child voices as there is not enough data to perform a large-scale evaluation such as a blizzard challenge. This objective evaluation was performed to see the correlation between reference child audio and synthetic child audio. A random set of 50 utterances were selected from the TinyMyST dataset as a part of this inside test. These utterances were used as reference utterances and the corresponding transcripts were used to generate synthetic speech for each of these utterances. This gave us 50 reference and 50 synthetic utterances which were used to calculate MOS using MOSNet. MOS score for 5 samples can be seen in Table 11. The spectrograms for these 5 samples can be seen in Figure 10.

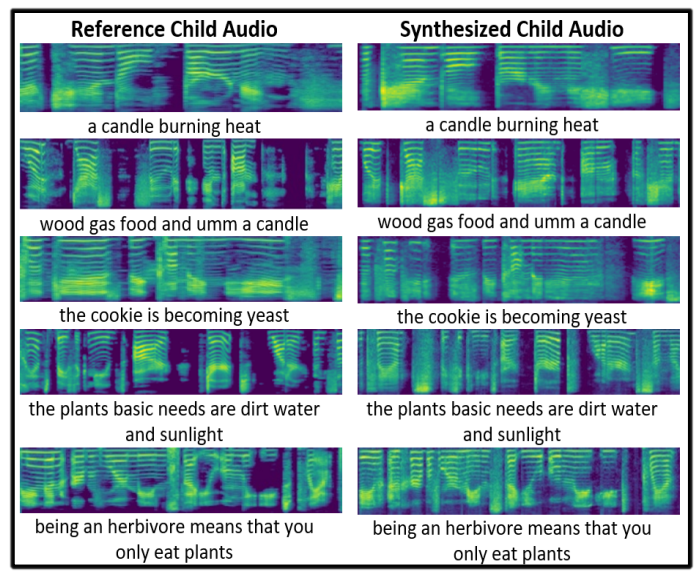

**FIGURE 10.** Spectrogram comparison between reference and synthesized child audio for 5 audio samples used with MOSNet.

**TABLE 11: MOSNET OUTPUT FOR 5 SAMPLES**

| Sample | Reference Child Speech MOS | Synthetic Child Speech MOS |
|--------|----------------------------|----------------------------|
| 1 | 2.25 | 2.80 |
| 2 | 3.08 | 2.77 |
| 3 | 3.18 | 2.91 |
| 4 | 3.17 | 2.51 |
| 5 | 3.17 | 2.99 |

**TABLE 12: MOSNET OUTPUT FOR 50 SAMPLES WITH 95% CONFIDENCE INTERVAL**

| Samples | Reference Child Audio MOS | Synthetic Child Audio MOS |
|---------|---------------------------|---------------------------|
| 50 | 2.91 ± 0.07 | 2.60 ± 0.06 |

Table 12 shows the overall MOS output for MOSNet. MOS of **2.96** was observed for **reference child audio** and **2.66** for **synthetic child audio**. There is only a 0.3 difference in MOS between reference and synthetic child voices. MOSNet trained on adult speech data is not expected to give MOS ratings correlated with human MOS ratings for child speech data. MOSNet was only used to get a correlation between the reference child audio and synthetic child audio. This gave us a comparison between reference and synthetic child voices as to how close they are to each other in terms of audio features calculated using MOSNet. The results confirmed that MOSNet output for reference child speech and synthetic child speech are very close to each other with a **comparative MOS difference** of **0.3**.

### C. SPEAKER SIMILARITY EVALUATION USING A SPEAKER VERIFICATION SYSTEM

Speaker similarity between a synthesized speech and a real speech can be calculated using a Speaker Verification (SV) system. The pretrained speaker encoder from section 2.B.1. was used with a third-party tool[10] to extract and visualize the speaker embeddings. This tool uses cosine distance to calculate the similarity between the two embeddings. The same speakers mentioned in our subjective evaluation (see Table 8) were used for this evaluation. 10 utterances were randomly selected for both real and synthetic speech for each of the 4 speakers mentioned in Table 8. 1 male and 1 female speaker from the LibriSpeech dataset were also added with 10 utterances each to show the speaker similarity comparison between an adult and child speaker. A visualization of this similarity in a 2D projection can be seen in Figure 11, 'gt' is used as a label for the ground truth of the speaker and 'ss' is used as a label for the synthetic speech of the same speaker. 'Adult_Male' and 'Adult_Female' are two randomly selected male and female speakers from the LibriSpeech Dataset.

From Figure 11, it can be inferred that Male, Female, and Child speech have a difference in similarity from each other. Male and Female adult speakers are far apart from each other and from child speakers in this 2D projection of speaker embeddings.

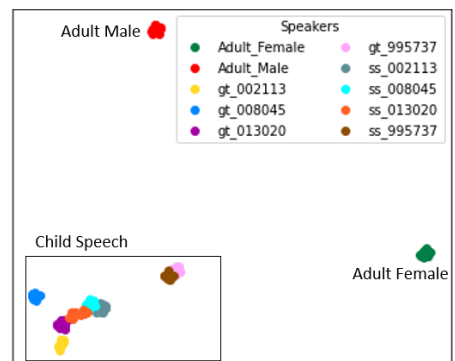

**FIGURE 11** Projections of embeddings between different real and synthetic child speech along with adult speech. The child Speech region [both ground truth and synthetic speech] is outlined by a solid black rectangle. The projections include a cluster of 10 voices selected from 10 different speakers. 'ss' refers to synthetic child speech and 'gt' refers to ground truth child speech.

To further comment on the similarity between real child speech and synthetic child speech, the 'child speech' contour from Figure 11 is extended to get a more visual representation of embeddings. This can be seen in Figure 12. The 'gt' labels and very close to 'ss' labels in this 2D projection space.

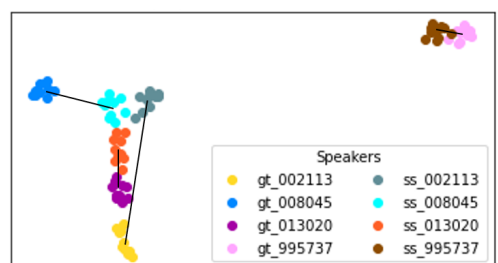

**FIGURE 12** Projections of embeddings between different real and synthetic child speech. A solid black line is used to show the distance between the ground truth and synthetic speech from the same speakers. This line was drawn from the centroid of each cluster to show the visual representation of similarity between real and synthetic speech.

These embeddings are 256-dimensional feature vectors trained by our speaker encoder. Therefore, cosine similarity was used to further calculate the cross-similarity between each speaker. Each of the 10 Speakers with 10 utterances each (1 Adult Male, 1 Adult Female, 4 Ground Truth Child, and 4 Synthetic Speech Child) were divided into 2 sets A and B. Embeddings are extracted for each of the utterances for each of the sets and averaged together for each speaker. This gave us 10 unique speaker embeddings in sets A and B each for 10 speakers. Cosine similarity is finally used to measure the similarity between each of the 10 speaker embeddings in sets A and B. A plot for the cross similarity between speakers can be seen in Figure 12.

---

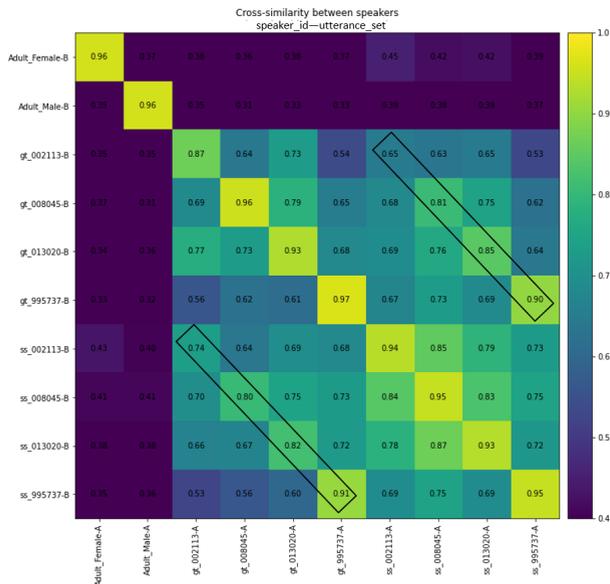

**FIGURE 13:** Cross-similarity between 10 speakers in Set A and Set B. The rectangular black box represents the similarity between real and synthetic child speech for respective speakers in set-A and set-B. Set-A is along the x-axis and Set-B is along the y-axis. 'ss' represents the synthetic speech and 'gt' represents the ground truth (real) speech.

In Figure 13, speaker similarity between synthetic speech and ground truth for speaker '995737' is 0.91, whereas for speakers '013020' and '008045' is approximately around 0.82 and finally for the speaker '002113' is approximately 0.7. This cross-similarity matrix gives us an idea of how close synthetic child voices are in comparison to the real child voice. It also shows us how different an Adult Male and Female Speech is in comparison to a child's speech. Overall, the similarity between most of the child and adult speech is between 0.3-0.4 whereas the similarity between most of the synthetic child speech and ground truth child speech is between a range of 0.65-0.85. Cross-similarity across the diagonal signifies that an utterance in set-A is 95% similar to utterances in set-B having the same speakers. More conclusions can be drawn from Figure 13, however, for the scope of this research, it is only used to show the different speaker similarities between real child speech and synthetic child speech and to draw a conclusion that our synthetically generated child speech is very close to real speech in terms of speaker similarity with an average similarity of 81%.

### D. OBJECTIVE INTELLIGIBILITY EVALUATION USING A PRETRAINED ASR SYSTEM

A pretrained wav2vec2 model is used to provide verification on synthetic utterances. A comparison of the speech transcription between real and synthetic child voices is presented. Child speech recognition is a challenging task of its own. The ASR on child speech is a part of our future work. Our intent to use this model for this paper is based on the popularity of the model, being SOTA on adult speech. A wav2vec2 model trained on adult speech data is used to provide that comparison. This speech transcription was obtained for the synthetic and real utterances mentioned in

Table 8. A random set of 30 utterances for each speaker for both real and synthetic voices are selected. The instruction for using this model is mentioned in their Github[11].

A comparison of this model is also provided using adult speech by selecting the equal number of adult voices from the LibriSpeech dataset. Word Error Rate (WER) is calculated from the output of wav2vec2 and is mentioned in Table 13. The Flashlight[12] library is used to calculate the WER using Viterbi decoding. No external language model (LM) was used.

**TABLE 13: WER ON ADULT SPEECH, REAL CHILD SPEECH AND SYNTHETIC CHILD SPEECH**

| Data type | # of Utterances | WER |
|---|---|---|
| Adult Speech [LibriSpeech_test_clean] | 120 | 3.43 |
| Real Child Speech [From MyST] | 120 | 15.27 |
| Synthetic Child Speech [Based on MyST] | 120 | 25.63 |

From Table 13, it can be inferred that the WER for Adult Speech (Librispeech_test_clean) is 3.43, evidently, due to the model being trained on adult speech data, the WER for real child speech is 15.27 and in comparison, WER for synthetic utterances is 25.63. An ASR model was able to recognize 75% of the synthetic speech with a relative difference of 10 WER when compared with real child speech recognized by the same model for the same speakers.

## IV. CONCLUSION AND FUTURE WORK

In this paper, a pipeline for generating synthetic child speech in a limited training data scenario is proposed. A small set of child speech data is created by cleaning an existing child speech dataset and making it suitable for TTS training. A transfer learning approach is used to train our model with adult speech data in a pretraining setting and child speech data as low as 19 hours for fine-tuning. MOSNet based objective evaluation shows a high correlation between real and synthesized child voices. A subjective evaluation method suitable for synthesized child speech is also proposed and demonstrated. Subjective MOS of synthesized voices is observed as 3.95 for speech intelligibility, 3.89 for voice naturalness, and 3.96 for overall voice consistency which is very close to Natural speech MOS. These MOS values tell us about how good the synthesized child voices are. However, voice inconsistency for 'End of phrase quality' containing noise and unintelligible information was also observed. There is scope for improvement for these phrases. WER for synthetic child voices using a pretrained adult speech wav2vec2 ASR model came to be 25.63 as compared to WER of real child voices of 15.27. Synthetic child speech samples can be viewed in our GitHub repository[13]. Multi-speaker TTS can be the key to child speech synthesis with limited training data. Child speakers with speech duration between 5-7 minutes in TTS training gave 81% average cosine similarity with a synthetic speech from the same speakers. This choice of the model allows the TTS to learn useful speaker information which can be leveraged to produce better quality synthetic voices even with limited child speech.

---

[11] wav2vec2: https://github.com/pytorch/fairseq
[12] Flashlight: https://github.com/flashlight/flashlight
[13] https://c3imaging.github.io/ChildTTS/

For future work, our aim is to improve this method by incorporating more information to our multi-speaker TTS model such as duration predictor and energy as implemented in FastSpeech2 [12]. The trained vocoder was also finetuned on the TinyMyST dataset. However, there was no significant improvement in the quality of the generated audio waveforms and an additional noise was observed in some of the synthesis. More child speech data would be required to achieve any significant improvement over the quality of the vocoder. It is also intended to implement GAN-based SOTA Vocoders such as HiFi-GAN [19] for future experiments. More experiments such as training a forced aligner using children's voices is also part of our future work. It will help to generate more meaningful alignments for splitting the longer audio files to increase the training dataset. The information collected from our subjective evaluation such as voice consistency in Harvard sentences will be used to improve child speech. This information will be used to collect better TTS-based child speech data based on Harvard sentences to accord with 'end of the phrase' information loss and voice inconsistency observed with our current results. The use of synthetically generated child speech to improve other areas of child speech research such as ASR and speaker recognition will also be investigated in future work. TTS-generated child voices can be used as a data augmentation technique for training these models with additional data. It is also intended to use the subjective evaluation method proposed in this paper for performing all future subjective evaluations with TTS generated child speech.

## ACKNOWLEDGMENT
The authors would like to acknowledge experts from Xperi-Ireland: Gabriel Costache, George Sterpu, and the rest of the team members for providing their expertise and feedback throughout.

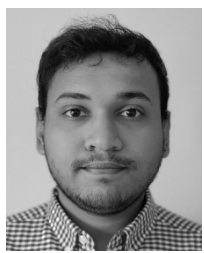

**RISHABH JAIN** received the B.Tech. degree in computer science and engineering from Vellore Institute of Technology (VIT), in 2019, and the M.S degree in data analytics from National University of Ireland Galway (NUIG), in 2020. He is currently working as a Research Assistant in NUIG under DAVID (Data-center Audio/Visual Intelligence on-Device) project. He is also pursuing the Ph.D. degree with the NUIG. His research interests include machine learning and artificial intelligence specifically in domain of speech understanding, text-to-speech, speaker recognition and automatic speech recognition.

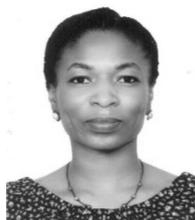

**MARIAM YAHAYAH YIWERE** received her Bachelor of Science degree from the Department of Computer Science at the Kwame Nkrumah University of Science and Technology in Kumasi, Ghana in 2012. She received her Master of Engineering degree and PhD degree from the department of Computer Engineering, Hanbat National University, South Korea in August 2015, and February 2020 respectively. Since October 2020, Mariam has been working on the DTIF/DAVID project as a postdoctoral researcher at the College of Science and Engineering, National University of Ireland, Galway. Her research interests include Text-to-Speech Synthesis, Speaker Recognition and Verification, Sound Source Localization, Deep Learning and Computer Vision.

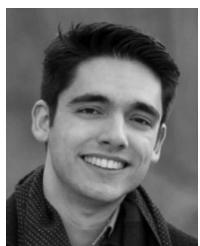

**DAN BIGIOI** graduated from the National University of Ireland Galway in 2020 with a bachelor's degree in Electronic and Computer Engineering. Upon graduating, he worked as a research assistant at NUIG studying the text to speech and speaker recognition methods under the DAVID (Data-Center Audio/Visual Intelligence on-Device) project. Currently, he is working on his Ph.D. at NUIG, sponsored by D-REAL, the SFI Centre for Research Training in Digitally Enhanced Reality. His research involves studying and implementing novel deep learning-based techniques for Automatic Speech Dubbing and discovering new ways to process multi-modal audio/visual data.

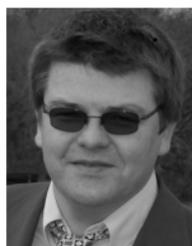

**PETER CORCORAN** (Fellow, IEEE) currently holds the Personal Chair in electronic engineering with the College of Science and Engineering, National University of Ireland Galway (NUIG). He was the Co-Founder of several start-up companies, notably FotoNation (currently the Imaging Division, Xperi Corporation). He has more than 600 cited technical publications and patents, more than 120 peer-reviewed journal articles, 160 international conference papers, and a co-inventor on more than 300 granted U.S. patents. He is an IEEE Fellow recognized for his contributions to digital camera technologies, notably in-camera red-eye correction and facial detection. He is also a member of the IEEE Consumer Technology Society for more than 25 years and the Founding Editor of IEEE Consumer Electronics Magazine.

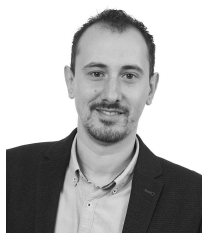

**HORIA CUCU** received the B.S. and M.S. degrees in applied electronics from University Politehnica of Bucharest (UPB), Romania, in 2008 and the Ph.D. degree in electronics and telecom from the same university in 2011.
From 2010 to 2017, he was a Teaching Assistant and then Lecturer at UPB. He currently serves as Associate Professor at the same university. In this position, he authored over 75 scientific papers in international conferences and journals, served as project director for 7 research projects, and contributed as a researcher to 10 other research grants. He holds two patents. In addition, he founded and leads Zevo Technology, a speech start-up dedicated to integrating state-of-the-art speech technologies in various commercial applications. His research interests include machine/ deep learning and artificial intelligence, with a special focus on automatic speech and speaker recognition, text-to-speech synthesis, and speech emotion recognition.
Dr. Cucu was awarded the Romanian Academy prize "Mihail Drăgănescu" (2016) for outstanding research contributions in Spoken Language Technology, after developing the first large-vocabulary automatic speech recognition system for the Romanian language.